\documentstyle[preprint,aps,psfig,tighten]{revtex}

\begin{document}
\draft
\title{Ginzburg-Landau equations for superconducting\\
quark matter in neutron stars}
\author{D.~Blaschke}
\address{Institute for Nuclear Theory, University of Washington\\
Box 351550, Seattle, WA 98195 \\
and\\
Fachbereich Physik, Universit\"at Rostock, D-18051 Rostock, Germany\\
and\\
Bogoliubov Laboratory for Theoretical Physics, Joint Institute for Nuclear\\
Research\\
14 19 80 Dubna, Russia}
\author{D.~Sedrakian}
\address{Observatoire de Paris, F-92195 Meudon Cedex, France \\
and\\
Department of Physics, Yerevan State University, 375 025 Yerevan, Armenia}
\maketitle

\begin{abstract}
We investigate magnetic properties of color superconducting quark matter
within a Ginzburg-Landau approach. The simultaneous coupling of the quark
fields to gluonic and electromagnetic gauge fields leads to {\it rotated
electromagnetism} with a massive (Higgsed) and a massless photon-gluon
field. We derive the Ginzburg-Landau equations for superconducting quark
matter taking into account the rotated electromagnetism in the general case
when the rotation angle $\alpha$ is an arbitrary function of the
coordinates. We solve these equations for the isolated vortex in
superconducting quark matter. We obtain a solution for the magnetic and
gluomagnetic fields and expressions for the calculation of the penetration
depth and the quantized magnetic flux in quark matter. For this case we have
demonstrated that the occurence of electric and color Meissner currents is a
consequence of the color superconducting state of quark matter. 
\vspace{5mm}
\end{abstract}


\preprint{\ \parbox[t]{45mm} 
{MPG-VT-UR 202/00} }

\section{Introduction}

\label{introsec}

Recently, the possible formation of diquark condensates in QCD at finite
density has been reinvestigated in a series of papers following Refs. \cite
{arw98,r+98}. It has been shown that in chiral quark models with a
nonperturbative 4-point interaction motivated from instantons \cite{cd98} or
nonperturbative gluon propagators \cite{br98,brs99} the anomalous quark pair
amplitudes in the color antitriplet channel can be very large: of the order
of $100~ {\rm MeV}$. Therefore, in two-flavor QCD, one expects this diquark
condensate to dominate the physics at densities beyond the
deconfinement/chiral restoration transition and below the critical
temperature ($\approx 50~ {\rm MeV} $) for the occurence of this 
{\it two-flavor color superconductivity} (2SC) phase. 
In a three-flavor theory it has been found 
\cite{arw99,sw99} that there can exist a color-flavor locked (CFL) phase for
not too large strange quark masses \cite{abr99} where color
superconductivity is complete in the sense that diquark condensation
produces a gap for quarks of all three colors and flavors, which is of the
same order of magnitude as that in the two-flavor case.

The high-density phases of QCD at low temperatures are most relevant for the
explanation of phenomena in rotating compact stars - pulsars. Conversely,
the physical properties of these objects (as far as they are measured) could
constrain our hypotheses about the state of matter at the extremes of
densities. In contrast to the situation for the cooling behaviour of compact
stars where the CFL phase is dramatically different from the 2SC phase \cite
{bkv}, we don't expect qualitative changes of the magnetic field structure
between these two phases. Consequently, we will restrict ourselves here to
the discussion of the simpler two-flavor theory first. The 2SC phase window 
is expected to occur at densities just above the deconfinement transition 
inside a compact star where the strange quark flavor is either still confined
\cite{gocke} or the self consistently determined strange quark masses are 
large enough to entail color-flavor unlocking \cite{buballa}. 

According to Bailin and Love \cite{bl} the magnetic field of pulsars should
be expelled from the superconducting interior of the star due to the
Meissner effect and decay subsequently within $\approx 10^4$ years. 
For their estimate, they used a perturbative gluon propagator which yielded
a very small pairing gap and they made the assumption of a homogeneous
magnetic field. Since both assumptions seem not to be valid in general, we
have performed in Ref. \cite{bss} a reinvestigation of the question 
with the result that 2SC quark matter is a type-II superconductor which can
form vortices in response to the external magnetic field thus behaving
similar to protons in the superfluid neutron phase of a neutron star \cite
{ss80,ss91}. The magnetic field in the quark core of the neutron star can
then exist for periods much longer than the spin-down age of a pulsar.

The authors of Ref. \cite{abr} have considered this question taking into
account the {\it rotated electromagnetism}. They came to the conclusion that
the magnetic field will live in the quark core sufficiently long although it
does not form the structure of quantum vortices because it obeys the
force-free Maxwell equation.

In the present work we derive the Ginzburg-Landau equations with account of
rotated electromagnetism and solve these equations for the example of an
isolated quantum vortex in quark matter which demonstrates that there is
always an electric and color Meissner current when the quark matter is
superconducting. This conclusion holds also for the case of a homogeneous
external magnetic field as we have shown in a separate paper \cite{sbsv}.

\section{Ginzburg-Landau Equations}
\label{gle}

In the paper \cite{bss} we have obtained the Ginzburg-Landau equations for
relativistic superconducting quarks, supposing that the superconducting
quark matter phase with $ud$ diquark pairing has $P_p=0$, where $p$ is the
color antitriplet index. We study the consequences of superconducting quark
cores in neutron stars for the magnetic field of pulsars. We find that
within recent nonperturbative approach to the effective quark interaction
the diquark condensate forms a type II superconductor whereas previously
quark matter was considered as a type I superconductor \cite{bl}. In both
cases the magnetic field which is generated in the superconducting hadronic
shell of superfluid neutrons and superconducting protons can penetrate into
the quark matter core since it is concentrated in proton vortex clusters
where the field strength exceeds the critical value. Recently, in the paper 
\cite{abr}, discussing the Meissner effect for color superconducting quark
matter, the authors introduce the $\hat Q$- charge generator 
\begin{equation}  \label{charge}
\hat Q = Q + \eta P_8~,
\end{equation}
where $Q$ is the conventional electromagnetic charge generator and $P_8$ is
associated with the one of the gluons in the representation of the quarks 
\begin{eqnarray}
Q&=&\mbox{diag}\left(\frac 2 3, -\frac 1 3, -\frac 1 3\right)~~ \mbox{in
flavor}~ u, d, s ~\mbox{space}~, \\
P_8&=&\frac{1}{\sqrt{3}}\mbox{diag}\left(1, 1, -2\right)~~ \mbox{in color}~
r, g, b ~\mbox{space}~.
\end{eqnarray}
The condition they requested was that the $\hat Q$- charge of all Cooper
pairs which form the condensate vanishes 
\begin{equation}  \label{cond}
\hat Q d_p = 0~.
\end{equation}
Applying this condition (instead of $P_8=0$) for the derivation of the
Ginzburg-Landau equations, we will see that in the superconducting quark
matter two different types of magnetic fields can exist simultaneously. The
penetration depths $\lambda_q$ for these two magnetic fields are completely
different. For one of them the penetration depth is finite and much shorter
than has been found in \cite{bss}, but for the other one it becomes
infinitely large.

The free energy density in the superconducting quark matter with $ud$
diquark pairing can be written in the following form \cite{bkv} 
\begin{eqnarray}  \label{fe}
f&=&f_n +a d_p d_p^* + \frac 1 2 \beta (d_p d_p^*)^2 + \gamma |\hat P d_p|^2
+ \frac{({\rm rot} \vec A)^2}{8 \pi} + \frac{({\rm rot} \vec G_8)^2}{8 \pi}~,
\end{eqnarray}
where $a=dn/dE~t$, $\beta=dn/dE~7\zeta(3)(\pi k_B T_c)^{-2}/8$, $%
\gamma=p_F^2~\beta/(6 \mu^2)$ and $t=(T-T_c)/T_c$ with $T_c$ being the
critical temperature, $p_F$ the quark Fermi momentum. In zeroth order with
respect to the coupling constant, $dn/dE=\mu p_F/\pi^2$. $|\hat P d_p|^2$ is
the kinetic term of the free energy density (\ref{fe}) which is given by 
\begin{equation}  \label{kin}
|\hat P d_p|^2=|(\vec \nabla + i e \vec A Q + i g \vec G_8 P_8) d_p|^2~.
\end{equation}
If we take into account the condition (\ref{cond}) and use (\ref{charge}) we
will have 
\begin{equation}  
\label{cond2}
Q d = - \eta P_8 d~.
\end{equation}
Inserting (\ref{cond2}) into (\ref{kin}) we will get 
\begin{equation}  
\label{kin2}
|\hat P d_p|^2=|(\vec \nabla + i e \eta \vec A P_8 + i g \vec G_8 P_8)
d_p|^2~.
\end{equation}
Following the paper \cite{abr}, let us introduce instead of the original
gauge fields $\vec A$ and $\vec G_8$ the linear combinations $\vec A_x$ and 
$\vec A_y$ 
\begin{eqnarray}  
\label{ax}
\vec A_x&=&\frac{- \eta e \vec A + g \vec G_8}{\sqrt{\eta^2 e^2 + g^2}} 
= - \sin \alpha \vec A + \cos \alpha \vec G_8~, \\
\vec A_y&=&\frac{g\vec A + \eta e \vec G_8}{\sqrt{\eta^2 e^2 + g^2}} 
= \cos \alpha \vec A + \sin \alpha \vec G_8~.  
\label{ay}
\end{eqnarray}
This corresponds to a rotation with the angle $\alpha$ in the orthogonal
basis of the fields $\vec A_x$ and $\vec A_y$, where 
\begin{eqnarray}
\cos \alpha =\frac{g}{\sqrt{\eta^2 e^2 + g^2}}~,  \label{angle}
\end{eqnarray}
while the normalization is preserved 
\begin{equation}  \label{cond3}
\vec A_x^2 + \vec A_y^2 = \vec A^2 + \vec G_8^2 ~.
\end{equation}

At neutron star densities the gluons are strongly coupled ($g^2/4 \pi
\approx 1$) and the photons are of course weakly coupled ($e^2/4 \pi \approx
1/137$), so that $\alpha \approx \eta e /g$ is small. For the diquark
condensate, where blue-green and green-blue $ud$ quarks are paired, 
$\eta=1/\sqrt{3}$ and therefore $\alpha \approx 1/20$. 
Taking into account (\ref{ax}) and (\ref{ay}), the kinetic term (\ref{kin2}) 
will become 
\begin{equation}  \label{kin3}
|\hat P d_p|^2=(\vec \nabla - i q \eta \vec A_x) d_p^* (\vec \nabla + i q
\eta \vec A_x) d_p~,
\end{equation}
where 
\begin{equation}  \label{charge2}
q=\sqrt{\eta^2 e^2 + g^2} P_8~,
\end{equation}
and $P_8=1/\sqrt{3}$. We see that the kinetic term has the conventional
expression but instead of the charge of the $ud$ diquark pair equal to $e/3$
we have the new charge $q$ which is much larger than $e/3$ (about 20 times).
The Ginzburg-Landau equations are obtained in the usual way. If we demand
that the variation of the free energy with respect to the parameters $d^*$, 
$\vec A_x$ and $\vec A_y$ have to be equal to zero, we will get 
\begin{eqnarray}  \label{fe2}
0&=&a d_p + \beta (d_p d_p^*)d_p + \gamma (i\vec \nabla + q A_x)^2 d_p~,
\end{eqnarray}
the equation of motion for the diquark condensate and for the gauge fields 
\begin{eqnarray}  \label{gle2a}
0&=&\sin \alpha~{\rm rot}~{\rm rot} \vec A - \cos \alpha~{\rm rot}~{\rm rot} 
\vec G_8 - 4 \pi i q \gamma [d \vec \nabla d^* - d^* \vec \nabla d] - 8 \pi
q^2 \gamma |d|^2 \vec A_x~, \\
0&=&\cos \alpha~ {\rm rot}~{\rm rot} \vec A + \sin \alpha ~{\rm rot}~
{\rm rot} \vec G_8 ~.
\label{gle2b}
\end{eqnarray}
These equations can be written in the following form 
\begin{eqnarray}  \label{gleq2a}
\lambda_q^2~ {\rm rot}~{\rm rot} \vec A + \sin^2 \alpha \vec A 
&=& i \frac{\sin \alpha}{2 q} \frac{(d_p \vec \nabla d_p^* 
- d_p^* \vec \nabla d_p)}{|d|^2} + \sin \alpha \cos \alpha \vec G_8 \\
\lambda_q^2~ {\rm rot}~{\rm rot} \vec G_8 + \cos^2 \alpha \vec G_8 
&=& -i \frac{\cos \alpha}{2 q} \frac{(d_p \vec \nabla d_p^* 
- d_p^* \vec \nabla d_p)}{|d|^2} 
+ \sin \alpha \cos \alpha \vec A~,  
\label{gleq2b}
\end{eqnarray}
where 
\begin{equation}  \label{lq}
\lambda_q^{-1}=\sqrt{8 \pi \gamma}|d| q = \sqrt{-\frac{4 q^2 t p_F^3}{3 \pi
\mu}}~.
\end{equation}
This is the system of Ginzburg-Landau equations for superconducting quark
matter which takes into account the ``rotated electromagnetism''. When we
define in the equations (\ref{gleq2a}) and (\ref{gleq2b}) the order
parameter in the following form 
\begin{equation}
d_p=|d_p| {\rm e}^{i\phi}~,
\end{equation}
where $\phi$ is the phase of the order parameter, then this system of
equations can be written in the following form 
\begin{eqnarray}  
\label{gleq3a}
\lambda_q^2~ {\rm rot}~{\rm rot} \vec A + \sin^2 \alpha \vec A &=& i 
\frac{\Phi_q~\sin \alpha}{2 \pi} \vec \nabla \phi 
+ \sin \alpha \cos \alpha \vec G_8 \\
\lambda_q^2~ {\rm rot}~{\rm rot} \vec G_8 + \cos^2 \alpha \vec G_8 
&=& -i \frac{\Phi_q~\cos \alpha}{2 \pi} \vec \nabla \phi 
+ \sin \alpha \cos \alpha \vec A  
\label{gleq3b}
\end{eqnarray}
Here $\Phi_q$ has the well known form 
\begin{equation}
\Phi_q=\frac{2 \pi \hbar c}{q}~.
\end{equation}
If we take the contour integral on both sides of equation (\ref{gleq3a})
where the contour $L$ is chosen as the equatorial ring which limits the
quark matter core region and take into account that 
\begin{eqnarray}
\oint_L {\rm rot~rot} \vec{A}~d\vec{l} &=& \oint_L {\rm rot}\vec{B}~d\vec{l}
=0~, \\
\oint_L \vec{G}_8~d\vec{l} &=&0~,
\end{eqnarray}
so that the electrical currents and the vector potential of the gluon field
on the surface of the quark core must vanish, then we get 
\begin{equation}  
\label{cond4}
\oint_L \vec{A}~d\vec{l} = \int_S {\rm rot} \vec{A}~d\vec{S} 
= \int_S \vec{B}~d\vec{S} = \frac{\Phi_q}{\sin \alpha} N~.
\end{equation}
This means that the flux of the magnetic field through an arbitrary surface
over the contour $L$ is equal to the sum of fluxes which are generated by 
$N$ vortices each of which has a flux ${\Phi_q}/{\sin \alpha}$~.

When we calculate this flux we obtain 
\begin{equation}
\Phi_q^\prime=\frac{6 \pi \hbar c}{e} = 6~\Phi_0~,
\end{equation}
where $\Phi_0=2 \times 10^{-7}$ G cm$^2$ is the magnetic flux quantum. In
order to obtain this result we have used equation (\ref{gleq3a}). It is easy
to see that the same result can be obtained using equation (\ref{gleq3b}),
as to be expected.

\section{Solution of Ginzburg-Landau equations for isolated vortex in quark
matter}

We now consider the solution of the Ginzburg-Landau equations (\ref{gleq3a}, 
\ref{gleq3b}) for an isolated vortex situated in homogeneous superconducting
quark matter. To this end it is sufficient to find the solution of these
equations without the vortex terms. The contribution of an isolated vortex
is taken into account by the condition (\ref{cond4}) where $N=1$.
Consequently we have to solve the following system of equations 
\begin{eqnarray}  \label{gleq4a}
\lambda_q^2~ {\rm rot}~{\rm rot} \vec A + \sin^2 \alpha ~\vec A &=& \sin
\alpha \cos \alpha ~\vec G_8 \\
\lambda_q^2~ {\rm rot}~{\rm rot} \vec G_8 + \cos^2 \alpha ~\vec G_8 &=& \sin
\alpha \cos \alpha ~\vec A ~.  \label{gleq4b}
\end{eqnarray}
From these equations it is easy to see that 
\begin{equation}  \label{29}
{\rm rot}~{\rm rot} \vec{G}_8 = -\cot \alpha ~{\rm rot}~{\rm rot} \vec{A}~.
\end{equation}
When we act on both sides of equation (\ref{gleq4a}) with the operator ${\rm %
rot~rot}$ and take into account that $\alpha= {\rm const}$, we obtain 
\begin{equation}  \label{30}
\lambda_q^2~ {\rm rot}~{\rm rot}~{\rm rot}~{\rm rot} \vec{A} 
+\sin^2 \alpha ~{\rm rot}~{\rm rot} \vec{A} 
=\sin \alpha~\cos \alpha ~{\rm rot}~{\rm rot} \vec{G}_8~.
\end{equation}
Inserting (\ref{29}) into (\ref{30}), we obtain the equation for the
determination of the electromagnetic vector potential 
\begin{equation}  \label{31}
\lambda_q^2~ {\rm rot}~{\rm rot}~{\rm rot}~{\rm rot} \vec{A} 
+ {\rm rot}~{\rm rot} \vec{A} = 0~.
\end{equation}
Let us denote 
\begin{equation}  \label{32}
{\rm rot}~{\rm rot} \vec{A} = \vec{M}~.
\end{equation}
Then, equation (31) takes the form 
\begin{equation}  \label{33}
\lambda_q^2~ {\rm rot}~{\rm rot} \vec{M} + \vec{M} = 0~.
\end{equation}
Consequently, instead of equation (\ref{31}) we can solve the system of
equations (\ref{32}) and (\ref{33}).

Let us consider the solution of equation (\ref{33}). Since our problem has
cylindric symmetry, the unknown function shall depend only on the coordinate 
$r$ denoting the distance to the vortex. It is easy to see that the vectors 
$\vec{M}$ and $\vec{A}$ have only azumutal components $\vec{M}_\varphi(r)$
and $\vec{A}_\varphi(r)$. Then equation (\ref{33}) takes the form 
\begin{equation}
\frac{d^2M_\varphi(r) }{dr^2}+ \frac{1}{r}\frac{dM_\varphi(r) }{dr}
-(\frac{1}{r^2}+\frac{1}{\lambda_q^2}) M_\varphi(r)=0~.  
\label{34}
\end{equation}
The general solution of this equation which tends to zero at infinity is
given by 
\begin{equation}  \label{35}
M_\varphi(r) = c ~ K_1(r/\lambda_q)~,
\end{equation}
where $K_1$ is the modified Bessel function of first kind. Next we solve the
equation (\ref{32}). It is easy to see that a special solution of this
equation is $\vec{A} = - \lambda_q^2 \vec{M}$, and the general solution
which vanishes at $r \to \infty$ will be $c_2/r$~. Finally for the vector
potential we obtain 
\begin{equation}  \label{36}
A_\varphi(r) = c_1 ~ K_1(r/\lambda_q) + c_2/r ~.
\end{equation}
The nonvanishing component of the magnetic field has $z$- direction and is
determined by the formula 
\begin{equation}  \label{37}
B_z(r) = \frac{1}{r} \frac{d}{d r} \left(r~ A_\phi (r) \right)~.
\end{equation}
Inserting (\ref{36}) into (\ref{37}) we finally obtain 
\begin{equation}  \label{38}
B_z(r) = c_1 ~ K_0(r/\lambda_q) ~,
\end{equation}
where $K_0$ is the modified Bessel function of zero kind. The constant $c_1$
is determined from the equation (\ref{cond4}) as 
\begin{equation}  \label{39}
c_1=\frac{\Phi_q^\prime}{2\pi\lambda_q^2} ~.
\end{equation}
Then we have 
\begin{equation}  \label{40}
B_z(r) = \frac{\Phi_q^\prime}{2\pi\lambda_q^2} ~ 
K_0\left(\frac {r}{\lambda_q}\right) ~.
\end{equation}
If we introduce the gluomagnetic field $\vec{B}_8={\rm rot} \vec{G}_8$~,
then we can determine it using equation (\ref{29}). The condition that the
magnetic fields $\vec{B}$ and $\vec{B}_8$ have to vanish at $r \to \infty$,
immediately determine the solution for $\vec{B}_8$ 
\begin{equation}  \label{41}
\vec{B}_8 = - \vec{B} \cot \alpha ~,
\end{equation}
where $\vec{B}$ is given by equation (\ref{40}).

As can be seen from the solution (\ref{40}) the field of the quantum vortex
in quark matter formally has the same form as the field of a vortex in an
ordinary superconductor. Whereas in the case of the ordinary superconductor
in the definition of the flux $\Phi $ and the penetration depth $\lambda $
enters the same charge of the Cooper pair (i.e. $2{\rm e}$, where ${\rm e}$
is the charge of the electron), in quark matter in the definition of 
$\Phi _q$ enters the charge of the quark Cooper pair ${\rm e}/3$, and 
therefore $\Phi _q^{\prime }=6\Phi _0$, and in the definition of the 
penetration depth $\lambda _q$ enters the charge $q$, which is about $20$ 
times larger than ${\rm e}/3$. Therefore, the penetration depth $\lambda _q$ 
which is
proportional to $q^{-1}$ is $20$ times smaller than that of hadronic matter.
It is obvious from equation (\ref{41}) that the magnetic quantum vortex is
accompanied by a gluomagnetic quantum vortex the center of which coincides
with that of the former, but its force lines have the opposite direction to
the magnetic ones. The flux of the gluomagnetic vortex is greater than that
of the magnetic vortex by a factor $\cot \alpha $.

Let us consider the equations for the rotated magnetic fields 
$\vec{B}_x={\rm rot}\vec{A}_x$ and $\vec{B}_y={\rm rot}\vec{A}_y$. 
which we obtain from the equations (\ref {gle2a}), (\ref {gle2b})
\begin{eqnarray}
\lambda _q^2~{\rm rot}~{\rm rot}\vec{B}_x+\vec{B}_x 
&=&\Phi _q\hat{e}\delta (\vec{r})~,  \label{42} \\
{\rm rot}\vec{B}_y &=&0~.
\end{eqnarray}
The energy of isolated quark vortex is given by
\begin{equation}
E=\frac 1{8\pi }\int [\vec{B}_y^2+\vec{B}_x^2
+(\lambda _q{\rm rot}\vec{B}_x)^2]~dV
\end{equation}
Note that from equations (\ref{ay}) and  (\ref{41}) follows that the
magnetic field $\vec{B}_y$ of the isolated quark vortex is equal to zero.
The magnetic field $\vec{B}_x$ will be found from the solution of
equation (\ref{42}), which coincides with the well known equation for the
magnetic field of an isolated vortex in ordinary superconductors. We can
use now these solutions and calculate the energy of an isolated quark vortex 
in the form 
\begin{equation}
E=\left(\frac{\Phi _q}{8\pi \lambda _q}\right)^2\ln \frac{\lambda _q}{\xi _q}
\end{equation}
As we see from this expression the energy of quark vortex depends
logarithmically on the renormalized charge $q$.

In concluding this section let us add the following: If the quark matter
occupies a spherical volume of radius $a$, and the vortex passes through the
center of this volume then the force lines of the magnetic vortex outside of
this volume appear to have the dipole form, whereas the force lines for the
gluomagnetic field have to be confined within this volume because of the
confinement condition $\vec{G}_8(a)=0$. Outside of the quark matter region
we have only the ordinary magnetic field. We also note that we can use these
results in the calculation of the distribution of magnetic and gluomagnetic
fields in a neutron star if its core consists of superconducting quark
matter.

\section{Conclusion}

\label{conc}

In this article we derived the Ginzburg-Landau equations with account for 
{\it rotated electromagnetism}. From the form of these equations written
for the physical fields $\vec B$ and $\vec B_8$ it is obvious that they are
coupled, i.e. the appearance of one of them leads to the immediate
appearance of the other. We have solved these equations for an isolated 
quantum vortex in quark matter and have shown that the centers of 
electromagnetic and gluomagnetic vortices coincide. 
The distribution around the continuation
of the center of the vortex in quark matter for the fields $\vec{B}$ and 
$\vec{B}_8$ has the same form as for ordinary superconductors. 
Nevertheless, the flux of the magnetic vortex is 6 times the magnetic flux 
quantum $\Phi_0$ and the penetration depth in quark matter is $20$ times less 
than that in hadronic matter. 
Only the force lines of the magnetic field can leave the
quark matter volume whereas the gluomagnetic field is confined therein. As
we have demonstrated in this paper for the case of an isolated vortex in
quark matter the occurence of electric and color Meissner currents is a
consequence of the color-superconducting state of quark matter. 

\section*{Acknowledgement}

We thank M. Alford, K. Rajagopal, R. Rapp, E. Shuryak, D. Voskresensky and
J. Wambach for discussions. We acknowledge financial support from the DAAD
for the exchange program between the Universities of Rostock and Yerevan and
from the ECT* Trento for our participation at the workshop: {\it Physics of
Neutron Star Interiors}. D.S. is grateful to the Observatoire de Paris at
Meudon for its hospitality during a research visit; D.B. thanks the
Institute for Nuclear Theory at the University of Washington for its
hospitality and the Department of Energy for partial support during the
programs INT-00-1: {\it QCD at Nonzero Baryon Density} and INT-01-2: 
{\it Neutron Stars}. 

\end{document}